\documentclass[conference]{IEEEtran}
\usepackage[letterpaper, left=0.65in, right=0.65in, bottom=1in, top=0.8in]{geometry}
\IEEEoverridecommandlockouts

\usepackage[acronym]{glossaries}
\newacronym{BS}{BS}{base station}
\newacronym{PS}{PS}{phase-shifter}
\newacronym{RL}{RL}{reinforcement learning}
\newacronym{AP}{AP}{analog precoder}
\newacronym{FC-HBF}{FC-HBF}{fully-connected HBF}
\newacronym{FSA-HBF}{FSA-HBF}{fixed subarray HBF}
\newacronym{DSA-HBF}{DSA-HBF}{dynamic subarray HBF}
\newacronym{BF}{BF}{beamforming}
\newacronym{UE}{UE}{user equipment}
\newacronym{AWGN}{AWGN}{additive white gaussian noise}
\newacronym{MIMO}{MIMO}{multiple-input multiple-output}
\newacronym{MISO}{MISO}{multiple-input single-output}
\newacronym{RF}{RF}{radio frequency}
\newacronym{RIS}{RIS}{reconfigurable intelligent surfaces}
\newacronym{IOT}{IOT}{internet-of-things}
\newacronym{CL}{CL}{convolutional layer}
\newacronym{FDD}{FDD}{frequency division duplex}
\newacronym{TDD}{TDD}{time division duplex}
\newacronym{CSI}{CSI}{channel state information}
\newacronym{DNN}{DNN}{deep neural network}
\newacronym{DP}{DP}{digital precoder}
\newacronym{DL}{DL}{deep learning}
\newacronym{SVD}{SVD}{singular-value decomposition}
\newacronym{CNN}{CNN}{convolution neural network}
\newacronym{FDP}{FDP}{fully digital precoder}
\newacronym{SE}{SE}{spectral efficiency}
\newacronym{OFDM}{OFDM}{orthogonal frequency division multiplexing}
\newacronym{OMP}{OMP}{orthogonal matching pursuit}
\newacronym{FL}{FL}{fully-connected layer}
\newacronym{HSHO}{HSHO}{Hybrid Structured Heuristic Optimization}
\newacronym{HBF}{HBF}{hybrid beamforming}
\newacronym{IA}{IA}{initial access}
\newacronym{mm-Wave}{mm-Wave}{millimeter wave}
\newacronym{mMIMO}{mMIMO}{massive MIMO}
\newacronym{SINR}{SINR}{signal-to-interference-noise ratio}
\newacronym{SNR}{SNR}{signal-to-noise ratio}
\newacronym{RSSI}{RSSI}{received signal strength indicator}
\newacronym{PZF}{PZF}{phase zero forcing}
\newacronym{PSO}{PSO}{particle swarm optimization}
\newacronym{ZF}{ZF}{zero forcing}
\newacronym{O-FDP}{O-FDP}{optimal fully digital precoder}
\newacronym{JT}{JT}{joint transmission}
\newacronym{CU}{CU}{central unit}
\newacronym{MSE}{MSE}{mean square error}
\newacronym{CEL}{CEL}{cross entropy loss}
\newacronym{CB}{CB}{conjugate beamforming}
\newacronym{NC}{NC}{network controller}
\newacronym{CoMP}{CoMP}{coordinated multi point}
\newacronym{CF-mMIMO}{CF-mMIMO}{cell-free massive MIMO}
\newacronym{CF-HBF}{CF-HBF}{cell-free hybrid beamforming}
\newacronym{CF-BF}{CF-BF}{cell-free beamforming}
\newacronym{MLDG}{MLDG}{meta-learning domain generalization}
\newacronym{MAML}{MAML}{model agnostic meta-learning}
\newacronym{WSR}{WSR}{weighted sum rate}
\newacronym{WMMSE}{WMMSE}{weighted minimum mean square error}
\newacronym{NN}{NN}{neural network}
\newacronym{LOS}{LOS}{line of sight}
\newacronym{NLOS}{NLOS}{non line of sight}
\newacronym{ML}{ML}{machine learning}
\newacronym{FCL}{FCL}{fully connected layer}
\newacronym{SSL}{SSL}{semi-supervised learning}
 
\newif\ifDeepMIMOModel
\DeepMIMOModeltrue

\newif\ifSimpleNParamEq
\SimpleNParamEqtrue

\usepackage{lettrine}
\usepackage{multirow}
\usepackage{longtable}
\usepackage[table,xcdraw]{xcolor}
\usepackage[linesnumbered,ruled,vlined]{algorithm2e}
\usepackage{float}
\makeatletter
\let\oldlt\longtable
\let\endoldlt\endlongtable
\def\longtable{\@ifnextchar[\longtable@i \longtable@ii}
\def\longtable@i[#1]{\begin{figure}[t]
\onecolumn
\begin{minipage}{0.5\textwidth}
\oldlt[#1]
}
\def\longtable@ii{\begin{figure}[t]
\onecolumn
\begin{minipage}{0.5\textwidth}
\oldlt
}
\def\endlongtable{\endoldlt
\end{minipage}
\twocolumn
\end{figure}}
\makeatother
\usepackage{ifpdf}
\ifCLASSINFOpdf
  \usepackage[pdftex]{graphicx}
  \usepackage{graphicx,epstopdf}
  \graphicspath{{../pdf/}{../jpeg/}}
  \DeclareGraphicsExtensions{.pdf,.jpeg,.png,.eps}
\else
  \usepackage[dvips]{graphicx}
  \usepackage{graphicx,epstopdf}
  \graphicspath{{../eps/}}
  \DeclareGraphicsExtensions{}
\fi
\usepackage{tikz}
\usetikzlibrary{decorations.pathreplacing,calc}
\newcommand{\tikzmark}[1]{\tikz[overlay,remember picture] \node (#1) {};}

\newcommand*{\AddNote}[4]{%
    \begin{tikzpicture}[overlay, remember picture]
        \draw [decoration={brace,amplitude=0.5em},decorate,line width=.2mm,black]
            ($(#3)!(#1.north)!($(#3)-(0,1)$)$) --  
            ($(#3)!(#2.south)!($(#3)-(0,1)$)$)
                node [align=center, text width=2.5cm, pos=0.5, anchor=west] {#4};
    \end{tikzpicture}
}%

\usepackage{cite}
\usepackage{amsthm}
\usepackage{steinmetz}
\usepackage{amssymb}
\usepackage{balance}
\usepackage{eqparbox}
\usepackage{multirow}
\usepackage{bbm}
\usepackage{float}
\SetAlFnt{\small}

\usepackage{amsmath}

\usepackage{mathtools, nccmath}

\usepackage{booktabs, siunitx}
\usepackage[scaled]{DejaVuSansMono}
\usepackage[T1]{fontenc}

\usepackage{color}
\usepackage{relsize}
\usepackage{mathtools}

\newcommand{\bs}[1]{\boldsymbol{#1}}

\newcommand{\mb}[1]{\mathbf{#1}}

\usepackage{mathtools}

\SetKwInput{KwInput}{Input}                
\SetKwInput{KwOutput}{Output}
\SetKwInput{KwOutputr}{Output Regression}              
\SetKwInput{KwOutputc}{Output Classification}              

\newcommand{\bseq}{\begin{subequations}}
\newcommand{\eseq}{\end{subequations}}
\newcommand{\baln}{\begin{align}}
\newcommand{\ealn}{\end{align}}
\newcommand{\balnd}{\begin{aligned}}
\newcommand{\ealnd}{\end{aligned}}
\newcommand{\beq}{\begin{equation}}
\newcommand{\eeq}{\end{equation}}
\newcommand{\beqn}{\begin{eqnarray}}
\newcommand{\eeqn}{\end{eqnarray}}
\newcommand{\beqno}{\begin{eqnarray*}}
\newcommand{\eeqno}{\end{eqnarray*}}
\newcommand{\bma}{\begin{displaymath}}
\newcommand{\ema}{\end{displaymath}}
\newcommand{\bnu}{\begin{enumerate}}
\newcommand{\enu}{\end{enumerate}}
\newcommand{\bce}{\begin{center}}
\newcommand{\ece}{\end{center}}
\newcommand{\btb}{\begin{tabular}}
\newcommand{\etb}{\end{tabular}}
\newcommand{\ba}{\begin{array}}
\newcommand{\ea}{\end{array}}
\usepackage{footnote}
\makesavenoteenv{tabular}
\makesavenoteenv{table}
\makeatletter 
\newcommand\semiHuge{\@setfontsize\semiHuge{21.1}{27.38}}
\makeatother

\usepackage{subfigure}
\usepackage{float}
\begin{document}
\title{A Low-Complexity Plug-and-Play Deep Learning Model for Massive MIMO Precoding Across Sites\\
\author{\IEEEauthorblockN{Ali~Hasanzadeh~Karkan$^{*}$, Ahmed~Ibrahim$^{\dagger}$, Jean-François~Frigon$^{*}$, and François~Leduc-Primeau$^{*}$\\
$^{*}$Department of Electrical Engineering, Polytechnique Montréal, Montréal, QC H3C 3A7, Canada\\
$^{*}$Emails: \{ali.hasanzadeh-karkan, j-f.frigon, francois.leduc-primeau\}@polymtl.ca. \\
$^{\dagger}$Ericsson Canada's R\&D,  Kanata, ON K2K 2V6, Canada, Email: ahmed.a.ibrahim@ericsson.com.
}
}
}

\maketitle
\IEEEpubidadjcol

\begin{abstract}
Massive multiple-input multiple-output (mMIMO) technology has transformed wireless communication by enhancing spectral efficiency and network capacity. This paper proposes a novel deep learning-based mMIMO precoder to tackle the complexity challenges of existing approaches, such as weighted minimum mean square error (WMMSE), while leveraging meta-learning domain generalization and a teacher-student architecture to improve generalization across diverse communication environments. 
When deployed to a previously unseen site, the proposed model
achieves excellent sum-rate performance while maintaining low computational complexity by avoiding matrix inversions and by using a simpler neural network structure.
The model is trained and tested on a custom ray-tracing dataset composed of several base station locations. The experimental results indicate that our method effectively balances computational efficiency with high sum-rate performance while showcasing strong generalization performance in unseen environments.
Furthermore, with fine-tuning, the proposed model outperforms WMMSE across all tested sites and SNR conditions while reducing complexity by at least 73$\times$.
\end{abstract}


\section{Introduction} \label{Sec:Intro}
\Gls{mMIMO} technology has ushered in a transformative era in wireless communication, offering remarkable spectral efficiency and network capacity improvements.
Downlink precoding is a crucial technique for enhancing the performance of multi-antenna cellular networks, particularly in \gls{mMIMO} systems. The optimization of precoding vectors is often formulated as a sum-rate maximization problem, which is non-convex and NP-hard \cite{luo2008dynamic}. This problem is typically addressed using iterative methods, such as the \gls{WMMSE} algorithm, known for its effectiveness in many scenarios \cite{shi2011iteratively}. 

In addition to the above optimization methods, some solutions leverage \gls{DL} techniques to simplify the precoder design. The authors in \cite{hojatian2021unsupervised} train a self-supervised model to predict the precoding matrix. In contrast, \cite{9403959} proposes unfolding the WMMSE algorithm by building a model that learns a specific optimization procedure for the WMMSE parameters, following the original WMMSE solution. Moreover, the works in \cite{yang2022learning, 8935405, lyu2023downlink} take a different approach by predicting individual components of the \gls{WMMSE} algorithm, which are combined to reconstruct the precoding matrix.

However, solving the sum-rate maximization problem using the \gls{WMMSE} algorithm requires iterative optimization and matrix inversions, which significantly increase the computational load and delay, particularly as the number of antennas increases, leading to an order of complexity of $O(N^3_{\sf{T}})$ \cite{shi2011iteratively}. Instead, approaches that leverage deep learning to directly estimate the precoding matrix can achieve better tradeoffs between performance and complexity \cite{hojatian2021unsupervised}. Nonetheless, these methods are limited in their ability to generalize when the \gls{CSI} distribution changes. Therefore, while \gls{DL}-based techniques offer the advantage of reduced complexity and faster inference, their effectiveness is constrained by their lack of adaptability to changing channel conditions. 

In this paper, we propose a ``plug-and-play precoder'' (PaPP) estimator for \gls{mMIMO} communication systems. This approach leverages \gls{WMMSE} for training while incorporating \gls{DL} techniques. The PaPP maintains high performance in terms of sum-rate, even with changes in base station locations. To ensure robust performance across different deployment environments, we incorporate \gls{MLDG} in the training, which enhances the model's ability to generalize effectively.
To reduce complexity, we employ a teacher-student training approach \cite{hu2023teacher}. This approach leverages a well-performing, complex model (the teacher) to transfer knowledge to a simpler, more efficient model (the student). Here, the student model learns to approximate the teacher’s predictions, enabling it to retain much of the teacher’s accuracy while operating with significantly reduced computational complexity. Additionally, we reduced training complexity by training the teacher model using a \gls{SSL} approach for precoding matrix estimation. This setup is ideal for field deployment, where the lightweight and faster student model can be used in real-time applications with limited computational resources.
Given the computational intensity of matrix inversion in the WMMSE algorithm, we implement simple \glspl{FCL} in the student model to estimate the precoding matrix based on the teacher model’s WMMSE-driven predictions.

Finally, to test the system in a variety of propagation environments, we assemble a novel dataset based on a realistic map of Montreal, incorporating buildings and obstacles to simulate urban environments. This allowed us to test the PaPP with a variety of base station configurations, demonstrating strong performance in diverse and previously unseen scenarios.

The remainder of this paper is structured as follows. Section \ref{Sec:Problem_Formulation} introduces the system model and problem formulation, outlining WMMSE for precoding in massive MIMO systems. Section \ref{Sec:Proposed} presents the PaPP, detailing the algorithm, architecture, and dataset construction. Section \ref{Sec:Complexity} compares the computational complexity and sum-rate performance of the PaPP with existing approaches. Section \ref{Sec:Simulation} discusses numerical results, highlighting the generalization capabilities and efficiency of the PaPP. Finally, Section \ref{Sec:conclusion} concludes the paper.

\section{Problem Formulation} \label{Sec:Problem_Formulation}
\subsection{System Model}
In this paper, we study a time division duplex multi-user \gls{mMIMO} system, where uplink channel estimates can be used to calculate the downlink precoder. This \gls{mMIMO} system has a \gls{BS} that is equipped with $N_{\sf{T}}$ antennas. The system is designed to concurrently support $N_{\sf{U}}$ users, each utilizing a single antenna. This configuration enables efficient communication by leveraging the multiple antennas at the \gls{BS} to enhance signal quality and increase capacity, ultimately allowing for simultaneous transmissions to multiple users. 

The received signal at the $k^{th}$ user can be expressed as
\begin{equation}\label{eq:signal_recived}
    \mathbf{y}_k =  \mathbf{h}_{k}^{\dagger} \sum_{\forall k}  \mathbf{w}_{k} x_k + \bs{\eta}_k \, ,
\end{equation}
where the wireless channel vector between the base station (\gls{BS}) and the $k^{th}$ user is represented by $\mathbf{h}_{k} \in \mathbb{C}^{N_{\sf{T}} \times 1}$. Let $x_k$
denote an independent transmitted complex symbol for a user. The term $\bs{\eta}_k \sim \mathcal{CN}(0, \sigma^2)$ indicates complex \gls{AWGN} characterized by a zero mean and a variance of $\sigma^2$. The downlink \gls{FDP} vector is defined as $\mathbf{W} = \left[ \mathbf{w}_1, \ldots, \mathbf{w}_k, \ldots, \mathbf{w}_{N_{\sf{U}}} \right] \in \mathbb{C}^{N_{\sf{T}} \times N_{\sf{U}}}$.
The associated \gls{SINR} user $k$ is given by
\begin{equation}
    \text{SINR}(\mb{w}_{k}) = \frac{ \big|\mb{h}^{\dagger}_{k} \mb{w}_{k} \big|^2}{\sum_{j \neq k} \big|\mb{h}^{\dagger}_{k} \mb{w}_{j} \big|^2 + \sigma^2}\,,
\end{equation}
and the sum rate for all users is
\begin{equation}\label{eq:sum-rate}
    R(\mb{W}) = \sum_{\forall k}  \text{log}_2 \Bigl(  1+ \text{SINR}(\mb{w}_{k}) \Bigr).
\end{equation}
The objective is to determine a precoding matrix $\mathbf{W}$ that maximizes the throughput in eq. (\ref{eq:sum-rate}) while adhering to the maximum transmit power constraint, $P_{\text{max}}$. Consequently, the downlink sum-rate maximization problem can be formulated as
\begin{equation}\label{eq:maximization}
\begin{aligned}
    & \underset{\mb{W}}{\max}~ R(\mb{W}) , \\
    \text{s.t.}  &\sum_{\forall k} \mb{w}_{k}^{\dagger}  \mb{w}_{k} \leq P_{\sf{max}} \, .
\end{aligned}
\end{equation}

\subsection{Weighted Minimum Mean Square Error Algorithm}
The sum-rate maximization problem in (\ref{eq:maximization}) is classified as NP-hard. To find good approximate solutions, the iterative \gls{WMMSE} algorithm \cite{shi2011iteratively} is commonly used, where the problem is transformed to a corresponding problem focused on minimizing the sum of \gls{MSE} under the independence assumption of $x_k$ and $\bs{\eta}_k$. 
Key parameters in this framework include the receiver gain $u_k$ and a positive user weight $v_k$, which are used to obtain the \gls{MSE} covariance matrix. 
The solution is obtained by iteratively solving convex subproblems to generate updates on the receiver gains, user weights, and beamforming matrix.

First $\mathbf{W}^{(0)}$ is initialized while satisfying the constraint in \eqref{eq:maximization},
then, at each iteration $i$, variables are updated as defined below until a stopping criterion is satisfied:
\begin{align}
    v^{(\text{i})}_{k} = \frac{ \sum_{j =1}^{N_{\sf{U}}} \big|\mb{h}^{\dagger}_{k} \mb{w}^{(\text{i}-1)}_{j} \big|^2 + \sigma^2}{\sum_{j \neq k}^{N_{\sf{U}}} \big|\mb{h}^{\dagger}_{k} \mb{w}^{(\text{i}-1)}_{j} \big|^2 + \sigma^2},
\end{align}
\begin{align}
    u^{(\text{i})}_{k} = \frac{ \mb{h}^{\dagger}_{k} \mb{w}^{(\text{i}-1)}_{j} }{\sum_{j=1}^{N_{\sf{U}}} \big|\mb{h}^{\dagger}_{k} \mb{w}^{(\text{i}-1)}_{j} \big|^2 + \sigma^2},
\end{align}
\begin{align}\label{eq:construct_beamforming_matrix}
    \mb{w}^{(\text{i}+1)}_{k} =  u^{(\text{i})}_{k} v^{(\text{i})}_{k}\mb{h}_{k}(\sum^{N_{\sf{U}}}_{j=1}   v^{(\text{i})}_{j} |u^{(\text{i})}_{j}|^2 \mb{h}_{j}\mb{h}_{j}^{\dagger} + \mu \mathbf{I})^{-1},
\end{align}
where $\mu \geq 0$ is a Lagrange multiplier.
In addition to the high complexity of solving \gls{WMMSE}, the inherent non-convexity of sum-rate maximization means that WMMSE is only guaranteed to converge to a local optimum, not necessarily the global solution.

\section{Proposed Method} \label{Sec:Proposed}
In this work, we propose a \gls{DL}-based precoder that takes \gls{CSI} as input and outputs a precoding matrix, designed to be adaptable to various deployment scenarios while prioritizing energy efficiency. 
To address computational challenges, we replace the costly matrix inversion in eq.~(\ref{eq:construct_beamforming_matrix}) with a \gls{DNN} trained to estimate the parameters of the precoding matrix. 
However, our experiments have shown that simple training procedures are unable to generate DNN models with the same level of performance and generalization as \gls{WMMSE}.
We address this limitation using a teacher-student training approach combined with \gls{MLDG} training to enhance generalization and reduce complexity.
This method enables efficient adaptation to new channel environments with minimal data requirements, making \gls{DL}-based \gls{mMIMO} systems more practical for real-world applications.

\subsection{Ray-Tracing Dataset}
In this work, we introduce a custom dataset designed to model channel characteristics in a \gls{mMIMO} system and evaluate the performance of the PaPP, focusing on its adaptability and efficiency in simulated real-world wireless environments. The dataset is generated using the MATLAB Ray-Tracing toolbox, simulating realistic propagation conditions with both \gls{LOS} and \gls{NLOS} components. 
Base stations are deployed in several locations in the greater Montreal area, utilizing OpenStreetMap (OSM) \cite{OpenStreetMap} data to incorporate real-world structures and materials. The propagation environment is configured to account for up to 10 reflections with no diffraction. This setup ensures the dataset accurately captures realistic multipath characteristics influenced by building structures, terrain, and other environmental factors. The training dataset encompasses diverse locations, including areas such as ``Université de Montréal'', ``Parc'', ``Rachel'', ``Cathcart'', ``Old Port'', ``Sherbrooke'', and ``Okapark'', ensuring the model generalizes effectively across different environments. In this paper, we exclude three deployment datasets from the training and use them to test the generalization of the methods: ``Ericsson'' features an industrial environment with 75\% \gls{LOS} users, ``Decarie'' is a residential area with a balanced \gls{LOS}/\gls{NLOS} mix, and ``Sainte-Catherine'', a downtown area, has 75\% \gls{NLOS} users, providing a rigorous test for model adaptability.

\subsection{Neural Network Architecture}
\begin{figure}[t]
    \centering
    \includegraphics[width=\linewidth]{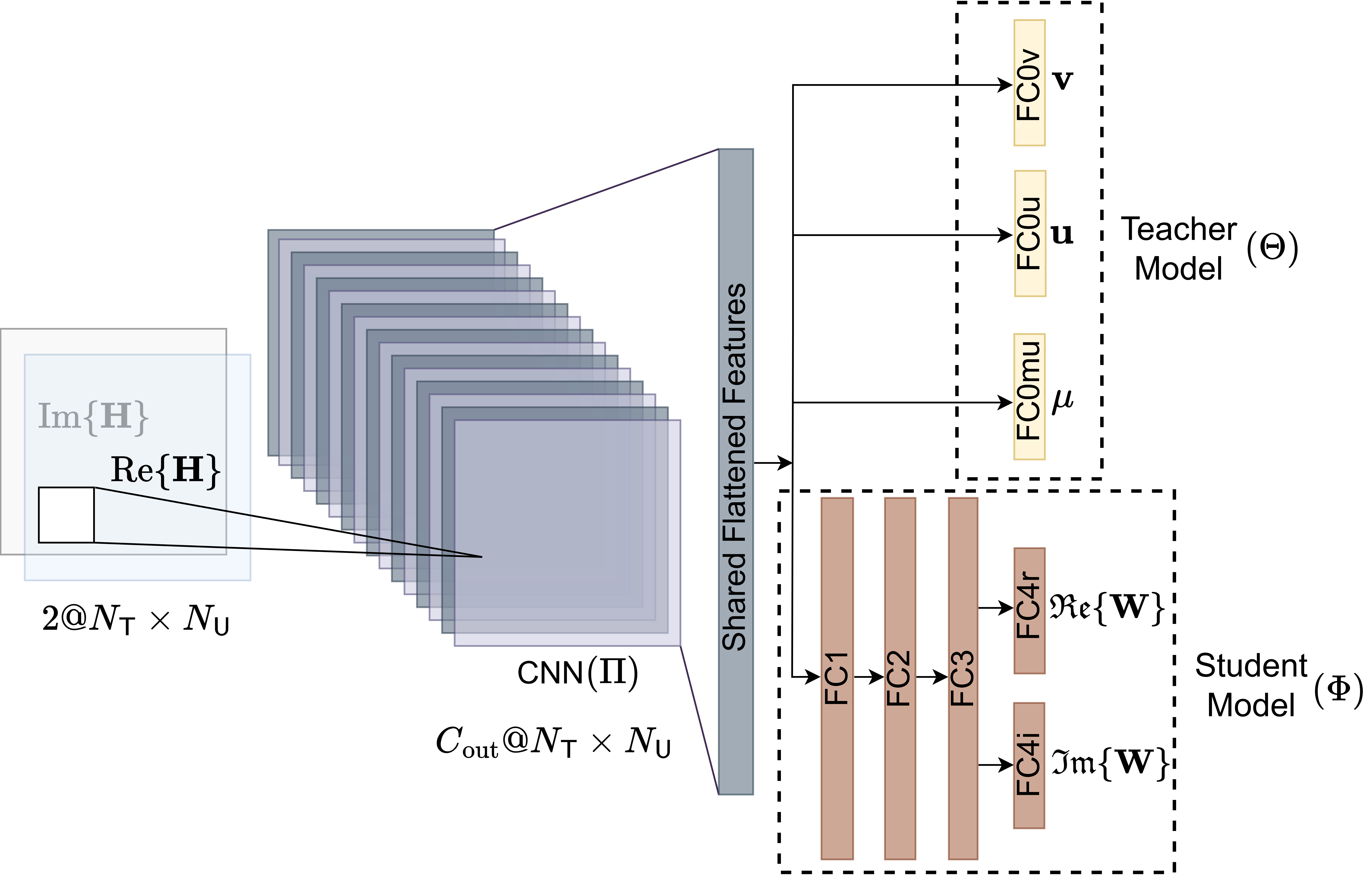}
    \caption{Proposed \gls{DNN} architecture.}
    \label{fig:arch}
\end{figure}
Our \gls{DNN} architecture for \gls{FDP} is illustrated in Figure \ref{fig:arch}. We employ a \gls{CNN} layer to extract features from the CSI, due to the larger number of antennas. The architecture is divided into three main components: \textit{Shared Feature Extraction}, \textit{Teacher Model}, and \textit{Student Model}. The teacher model contains three output layers to estimate $\mathbf{v}$, $\mathbf{u}$, and $\mu$, which are used to construct the precoding matrix using eq. (\ref{eq:construct_beamforming_matrix}).
The student model directly learns to estimate the precoding matrix under supervised training, guided by the constructed precoding matrix from the teacher model, while simultaneously maximizing the sum rate through \gls{SSL}. This architecture employs ReLU activation functions for non-linearity, along with 15\% dropout layers during backbone training to promote generalization. Batch normalization is also incorporated to improve convergence and stability, ensuring the model effectively adapts to \gls{mMIMO} environments while maintaining robust generalization performance.

\begin{algorithm}[t]
  \caption{PaPP Backbone Training}
  \label{Alg1}
  
  \KwIn{Different sites channel models $\bs{\mathcal{D}}$ (source domain)}
  \SetKwInOut{Init}{Initialize}
  \SetKw{IN}{in}
  \SetKw{each}{each}
  \newcommand{\forcond}{$e=1$ \KwTo $Epochs$}
  \newcommand{\forcondd}{$e=1$ \KwTo $Epochs$}
  \newcommand{\forcondom}{$i=1$ \KwTo $m$}
  \newcommand{\forcondou}{$j=1$ \KwTo $N_U$}
  \newcommand{\fordomains}{\each $\hat{D}$ \IN $\bs{\mathcal{D}}^{\textnormal{train}}$ }
  \newcommand{\fordomainsgen}{\each $\hat{D}$ \IN $\bs{\mathcal{D}}^{\textnormal{gen}}$ }
  \Init{Parameters $\Theta$ (Teacher), $\Phi$ (Student), $\Pi$ (Features)\\ 
  Learning rates $\alpha_{T}, \alpha_{S}, \alpha_{F}, \beta_{T}, \beta_{S}, \beta_{F}, \epsilon_{T}, \epsilon_{S}, \epsilon_{F}$}  
  
 \For{\forcond}{
  \textbf{Divide} $\bs{\mathcal{D}} \rightarrow \bs{\mathcal{D}}^{\textnormal{train}}$ and $\bs{\mathcal{D}}^{\textnormal{gen}}$~~~~~~~~~~~~~~~\tikzmark{right}\\
  $\delta_{T}, \delta_{S} \gets 0$\tikzmark{top}\\
  
  \For{\fordomains}{
  $\delta_{T} \leftarrow \delta_{T} + \frac{\nabla_{\Theta,\Pi}\mathcal{L}_{T}(\hat{D};\Theta,\Pi)}{|\bs{\mathcal{D}}^{\text{train}}|}$\\
  $\delta_{S} \leftarrow \delta_{S} + \frac{\nabla_{\Phi}\mathcal{L}_{S}(\hat{D}, \mb{W}_{T};\Phi)}{|\bs{\mathcal{D}}^{\text{train}}|}$\\
    }
  \textbf{Update} \\ 
    \qquad $\Theta^{\prime} = \Theta - \alpha_{T} \delta_{T}$\label{line:8}\\
    \qquad $\Pi^{\prime} = \Pi - \alpha_{F} \delta_{T}$\label{line:9}\\
    \qquad $\Phi^{\prime} = \Phi - \alpha_{S} \delta_{S}$ \label{line:10}\tikzmark{bottom}\AddNote{top}{bottom}{right}{Meta-Training}\\
  $\delta^{\prime}_{T}, \delta^{\prime}_{S} \gets 0$\tikzmark{top}\\
  \For{\fordomainsgen}{
  $\delta_{T}^{\prime} \leftarrow \delta_{T}^{\prime} + \frac{\nabla_{\Theta^{\prime},\Pi^{\prime}}\mathcal{L}_{T}(\hat{D};\Theta^{\prime},\Pi^{\prime})}{|\bs{\mathcal{D}}^{\text{gen}}|}$\\
  $\delta_{S}^{\prime} \leftarrow \delta_{S}^{\prime} + \frac{\nabla_{\Phi^{\prime}}\mathcal{L}_{S}(\hat{D}, \mb{W}^{\prime}_{T};\Phi^{\prime})}{|\bs{\mathcal{D}}^{\text{gen}}|}$\tikzmark{bottom}\\
    }
    \textbf{Meta-update:}\\ 
    \qquad $\Theta \leftarrow \Theta -  \epsilon_{T}(\delta_{T} + \beta_{T} \delta_{T}^{\prime})$\label{line:16}\\
    \qquad $\Pi \leftarrow \Pi -  \epsilon_{F}(\delta_{T} + \beta_{F} \delta_{T}^{\prime})$\label{line:17}\\
    \qquad $\Phi \leftarrow \Phi -  \epsilon_{S}(\delta_{S} + \beta_{S} \delta_{S}^{\prime})$\label{line:18}\AddNote{top}{bottom}{right}{Meta-Testing}
    \\
}
  \end{algorithm}
\subsection{Backbone Training}
The backbone model represents the generalized precoding model, trained for efficient deployment in diverse wireless environments. The backbone training process consists of three components: a teacher-student approach for knowledge transfer, \gls{SSL} to optimize the model, and \gls{MLDG} to enhance generalization.
We summarize the PaPP algorithm in Alg.~\ref{Alg1}. 

To enhance the adaptability of the precoding solution, we incorporated \gls{MLDG} as described in \cite{10624768}. The \gls{MLDG} framework is utilized during the backbone training phase, where the model is trained on channel data from different sites to improve both in-domain performance and out-of-domain adaptability. During this meta-training process, the model is exposed to a set of training domains and a separate set of generalization domains, ensuring it can effectively generalize to previously unseen distributions during deployment. At each iteration, we randomly divide the source domain $\bs{\mathcal{D}}$ which is a set of different site datasets to construct $\bs{\mathcal{D}}^{\text{train}}$ and $\bs{\mathcal{D}}^{\text{gen}}$. This two-phase training process (meta-training and meta-testing) enables the model to generalize better in various wireless channel environments. 

By employing a teacher-student training approach \cite{hu2023teacher}, we leverage a well-performing, complex model (the teacher with parameters $\Theta$) to transfer knowledge to a simpler, more efficient model (the student with parameters $\Phi$). In this approach, the student model learns to approximate the teacher's predictions, enabling it to perform effectively with significantly reduced computational complexity. 
The teacher precoder $W_T$ is obtained by performing a single iteration of the WMMSE, but using intermediate variables generated by the teacher DNN. Re-using (\ref{eq:construct_beamforming_matrix}), we set $\mathbf{W}_T = [\mathbf{w}_1^{(1)}, \dots, \mathbf{w}_k^{(1)}, \dots, \mathbf{w}_{N_{\sf{U}}}^{(1)}]$, while $u_k^{(0)}$, $v_k^{(0)}$, $\mu$ are obtained from the teacher DNN output. 
The student model uses shared features to directly predict the precoding matrix. Lines \ref{line:8}--\ref{line:10} in Alg.~\ref{Alg1} outline the parameter update process, where \(\Theta\), \(\Pi\), and \(\Phi\) are updated for meta-testing via gradient descent on \(\bs{\mathcal{D}}^{\text{train}}\). The feature extractor is updated only during teacher training. Lines \ref{line:16}--\ref{line:18} describe the meta-update process, updating \(\Theta\), \(\Pi\), and \(\Phi\) to optimize both domains’ losses and ensure harmonious convergence.

The teacher model is trained in a self-supervised manner, using sum-rate maximization as the objective function. This training regime ensures that the teacher model is optimized to directly predict effective precoding vectors without needing extensive labeled data, thereby enhancing generalization ability while reducing training complexity. The loss function of the teacher model is given by
\begin{align}
    \mathcal{L}_{T} = - R(\mb{W}_{T}) \,.
\end{align}
For training the student model, we design a loss function that ensures the student not only mimics the teacher but also optimizes itself to achieve a higher sum-rate. The following is the loss function for the student model: 
\begin{align}
    \mathcal{L}_{S} = 
\begin{cases}
    \mathcal{L}_{\text{MSE}}                             &  \text{if $R(\mb{W}_{T}) < 0.8\times R_{\text{WMMSE}}$}, \\
    \mathcal{L}_{\text{MSE}} - \lambda  R(\mb{W}) &  \text{otherwise},
\end{cases}
\end{align}
where $R_{\text{WMMSE}}$ represents the sum rate obtained using \gls{WMMSE} \cite{shi2011iteratively}, and
\begin{align}
    \mathcal{L}_{\text{MSE}} = \frac{1}{N_{\sf{T}} N_{\sf{U}}} \sum_{i=1}^{N_{\sf{T}}} \sum_{j=1}^{N_{\sf{U}}} \left ( \mb{W}_{T,ij} - \mb{W}_{ij} \right)^2 \,.
\end{align}
Parameter $\lambda$ is a regularization parameter that balances the trade-off between imitating the teacher model’s predictions and maximizing the student model's sum-rate. When the student model’s performance $R(\mb{W}_{T})$ is below 80\% of $R_{\text{WMMSE}}$, the loss focuses purely on reducing MSE. Otherwise, the term $\lambda R(\mb{W})$ encourages the student model to improve the sum-rate further. 

Once the training is complete, the model uses only the student component along with the trained feature extractor, enabling rapid precoding design with significantly reduced computational complexity.

\subsection{Deployment}
The fine-tuning process enables our model to adapt and perform effectively in real-world deployment scenarios online. During the deployment phase, the teacher portion of the original architecture is discarded, and the model operates in a zero-shot (i.e., without fine-tuning) mode or can be fine-tuned using \gls{SSL} on the entire model with varying amounts of local data. Furthermore, this local data is augmented by generating permutations of the CSI samples \cite{10624768}. This flexibility ensures that the model can adapt to new, unseen environments with minimal data, while still improving its performance with additional fine-tuning.

\begin{figure*}[!th]
    \centering
    \subfigure[Sum rate achieved at an average SNR of 40\,dB.]{\includegraphics[width=\columnwidth]{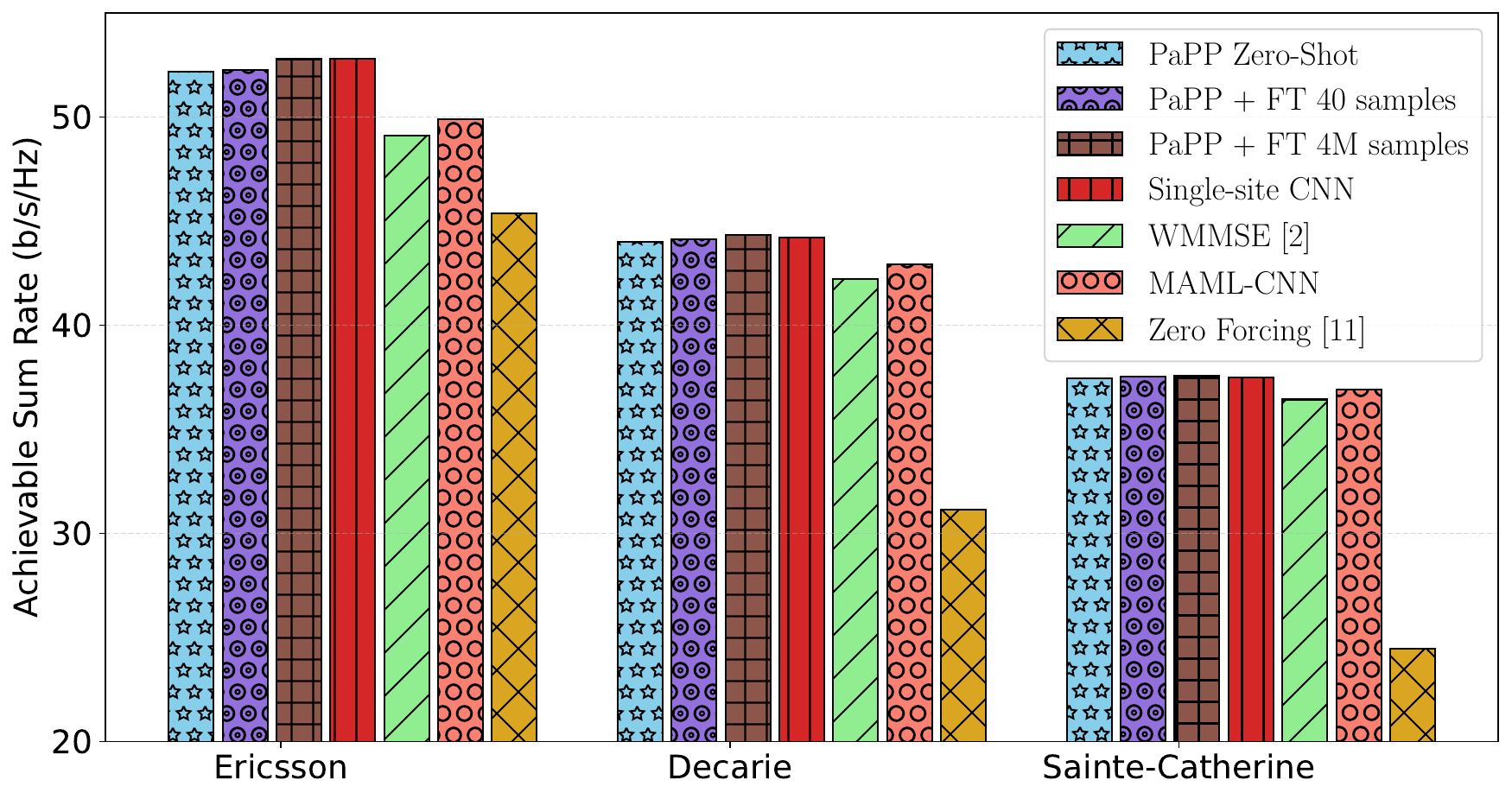}}
    \subfigure[Sum rate achieved at an average SNR of 10\,dB.]{\includegraphics[width=\columnwidth]{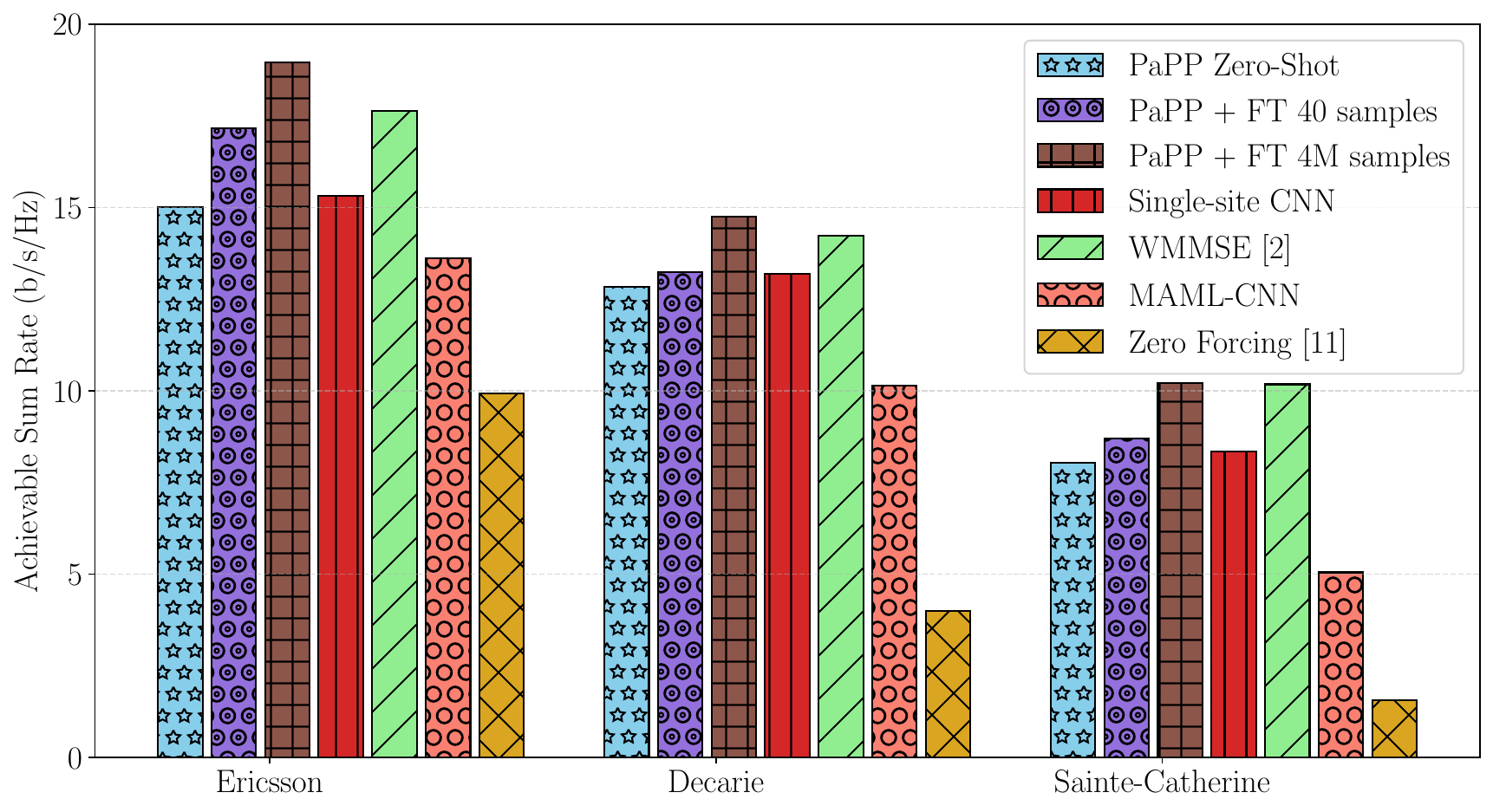}}
    \caption{Achievable sum rates for precoding methods under different SNR values at three Montreal sites: ``Ericsson'', ``Decarie'', and ``Sainte-Catherine''. All methods report zero-shot performance, except ``PaPP + FT,'' which is after 20 epochs of fine-tuning.}
    \label{fig:sidebyside}
\end{figure*}
\section{Baselines and Complexity Analysis} \label{Sec:Complexity}
\begin{table}[t]
    \centering
    \caption{Number of real multiplications and sum rate for different methods (``Ericsson'' site, $N_{\sf{T}} = 64$, $N_{\sf{U}} = 4$, SNR = 27dB).}
    \resizebox{\columnwidth}{!}{
    \begin{tabular}{ccc}
        \toprule
        \textbf{Methods} & \textbf{\# of Multiplications} & \textbf{Sum-Rate (b/s/Hz)}\\
        \midrule
        WMMSE \cite{shi2011iteratively} ($I=12.5$)    & 36.1\,M & 37.29
        \\ 
        MAML-CNN (zero-shot) & 3.77\,M & 37.01 \\
        \textbf{PaPP} (zero-shot) & \textbf{1.05\,M} & \textbf{38.10}\\
        Zero Forcing \cite{nayebi2017precoding} & 8.4\,K & 32.14\\
        \bottomrule
    \end{tabular} }
    \label{tab:complexity}
\end{table}
Table \ref{tab:complexity} shows an example of the computational complexity of different precoding methods, measured by the number of real multiplications required for processing when the deployment site is ``Ericsson''. This analysis offers insights into the trade-offs between computational demands and sum-rate performance.

The \gls{WMMSE} algorithm is renowned for achieving good weighted sum rate in mMIMO systems. However, this performance comes at the cost of high computational complexity. Specifically, with stopping criteria of $10^{-3}$ and an average of 12.5 iterations for this setup, the WMMSE method requires approximately 36 million multiplications. This complexity can limit real-time applications or systems constrained by computational resources. The total number of real multiplications for the WMMSE method is given by
\[
 4I \bigg(\frac{2}{3} N_{\sf{T}}^3N_{\sf{U}} + N_{\sf{T}}^2N_{\sf{U}} + 2N_{\sf{T}}(2N_{\sf{U}}^2+N_{\sf{U}})+ N_{\sf{U}}^2+\frac{14}{3}N_{\sf{U}}\bigg) \, ,
\]
where $I$ is the total number of iterations. 

As another baseline, we consider the method proposed in \cite{lyu2023downlink}, which combines a multilayer perceptron (MLP) model with the WMMSE algorithm. However, we replace the original MLP model with a CNN model similar to the approach in \cite{hojatian2021unsupervised}.
This modification was made to better capture spatial features in the \gls{CSI}, enabling the model to handle the increased complexity introduced by a larger number of antennas and users. By leveraging the representational power of CNNs, MAML-CNN achieves more competitive results in our experimental settings compared to the original MAML-MLP. Since it combines a \gls{DNN} with an additional matrix inversion step, it exhibits significantly higher complexity than \gls{ZF} but remains less demanding than WMMSE. 
The number of real multiplications required by MAML-CNN is given by
\begin{align*}
 & \
C_{\text{out}}N_{\sf{T}}N_{\sf{U}}C_{\text{in}}k^2 + C_{\text{out}}N_{\sf{T}}N_{\sf{U}}(3N_{\sf{U}}+1) \\
  & + 8\bigg(\frac{4}{3} N_{\sf{T}}^3 + N_{\sf{T}}^2 (3N_{\sf{U}}+2) + N_{\sf{T}}(2N_{\sf{U}}+3)\bigg) \, ,
\end{align*}
where $C_{\text{in}}$ and  $C_{\text{out}}$ are the input and output channels of the CNN layer, and $k$  is the kernel size. 

The proposed PaPP method leverages a DNN to provide a zero-shot precoding solution. While it exhibits greater computational complexity compared to traditional methods like ZF due to its convolutional and fully connected layers, this approach offers significant performance benefits, including improved interference mitigation and adaptability to unseen sites. The total number of real multiplications required for the PaPP method can be expressed as
\begin{align*}
 & \ C_{\text{out}}N_{\sf{T}}N_{\sf{U}}C_{\text{in}}k^2 + C_{\text{out}}N_{\sf{T}}N_{\sf{U}}D_{\text{FC1}} \\
& +D_{\text{FC1}}D_{\text{FC2}} + D_{\text{FC2}}D_{\text{FC3}} + D_{\text{FC3}}D_{\text{FC4r}} + D_{\text{FC3}}D_{\text{FC4i}}\, ,
\end{align*}
where $D_{\text{FCi}}$ ($i=1,2,3,4$) represent the sizes of the \glspl{FCL}. 

The \gls{ZF} \cite{nayebi2017precoding} precoding method offers significantly lower computational complexity, requiring approximately 8.4 thousand multiplications. This efficiency stems from its reliance on simpler linear algebra operations, specifically the inversion of smaller matrices when $N_{\sf{T}}>N_{\sf{U}}$. Despite its computational efficiency, ZF is susceptible to performance degradation in high-interference scenarios or under adverse channel conditions. The total number of real multiplications required by ZF is
\[
8 N_{\sf{U}}^2 N_{\sf{T}} + \frac{8}{3} N_{\sf{U}}^3 \, .
\]

\section{Numerical Results} 
\label{Sec:Simulation}
\subsection{Simulation Setup}
The detailed configuration parameters of the dataset and system setup are as follows. All base stations employ a uniform planar array with $N_{\sf{T}} = 8 \times 8$ elements, each spaced at half the wavelength $(\frac{\lambda}{2})$, operating at a carrier frequency of 2\,GHz. The transmitter antenna is positioned at a height of 20\,m, and the transmitter power is set to 20\,W. System losses are modeled with a 10\,dB margin to account for practical constraints in wireless communication. To generate the dataset, a circular arrangement of user locations is created around the BS, covering distances from 50\,m to 350\,m, and users spread evenly at 10 degree intervals. Each sample in the dataset includes $N_{\sf{U}}=4$ such user locations.

Table \ref{tab:hyper} summarizes the key hyperparameters used for training the PaPP model. These hyperparameters were carefully selected based on preliminary experimentation to achieve a balance between computational efficiency and model performance, ensuring robust operation across different deployment environments.
\begin{table}[t]
    \caption{Hyperparameter Settings For PaPP model Training}
    \centering
    \resizebox{\columnwidth}{!}{
    \begin{tabular}{cc}
        \toprule
        \textbf{Hyperparameter} & \textbf{Value}  \\
        \midrule
        Size of CNN output channels ($C_{\text{out}}$) & 32\\ 
        Size of FCLs (FC1, FC2, FC3) & 64, 64, 512 \\
        Regularization parameter in Loss function ($\lambda$) & 0.1\\
        Teacher model LRs ($\alpha_{T}, \beta_{T},\sigma_{T}$) & $10^{-1}, 10^{-2}, 10^{-2}$\\
        Feature model LRs ($\alpha_{F},\beta_{F},\sigma_{F}$) & $10^{-1}, 10^{-2}, 10^{-2}$\\
        Student model LRs ($\alpha_{S},\beta_{S},\sigma_{S}$) & $10^{-2}, 10^{-3}, 10^{-3}$\\
        Number of sites in each set $|\bs{\mathcal{D}}|, |\bs{\mathcal{D}}^{\text{train}}|, |\bs{\mathcal{D}}^{\text{gen}}|$ & 7, 5, 2\\
        Dataset size of each site $|\hat{\mathcal{D}}|$ & 1\,M\\
        Batch size & 1000\\
        \bottomrule
    \end{tabular}}
    \label{tab:hyper}
\end{table}

\subsection{Results}
Our experimental results show that \gls{DL}-based algorithms trained on a single site are sensitive to changes in the \gls{BS} location, resulting in performance degradation when deployed in new environments. For example, using the \gls{DL}-based method introduced in \cite{hojatian2021unsupervised}, if a precoder design model is trained in a downtown Montreal site (Old Port) and deployed to another downtown site (Sainte-Catherine), the zero-shot performance degrades by 53\% compared to when the model is trained on the same site. 

We analyze the performance of the PaPP model in comparison with \gls{ZF} \cite{nayebi2017precoding}, \gls{WMMSE} \cite{shi2011iteratively} (with stopping criteria of $10^{-3}$ and maximum number of iteration of 100), the MAML-CNN, and the single-site CNN, which is trained from scratch on the deployment site. The results are shown in Fig.~\ref{fig:sidebyside}.

At a high SNR (40\,dB), the PaPP zero-shot method consistently delivers competitive sum rates across all three locations, outperforming the WMMSE method by a margin of approximately 3--7\% and the MAML-CNN method by around 2--5\%. This suggests that while MAML-CNN leverages meta-learning principles, limitations prevent it from fully optimizing at high SNR conditions. Additionally, the PaPP trails the Single-site CNN approach by a negligible margin (<1\%). This demonstrates the outstanding adaptability of PaPP zero-shot to new sites without fine-tuning.
Meanwhile, the Z approach is consistently outperformed by the other methods, delivering rates that are approximately 15--53\% lower than the PaPP zero-shot.

 At a low SNR (10\,dB), the PaPP zero-shot method outperforms MAML-CNN by around 10--60\%, but trails WMMSE by 15--21\%, depending on the location. This suggests that while PaPP zero-shot leverages its design effectively, it requires fine-tuning on the deployment site to unlock its full potential under low-SNR conditions. The WMMSE iterative optimization outperforms all other zero-shot techniques across all sites. This highlights WMMSE's robustness in managing challenging noise-dominated scenarios.
 The PaPP zero-shot falls just 2--4\% behind the single-site CNN approach, which is a significant achievement considering it does so without any local data. This demonstrates the model's ability to effectively perform in diverse environments with zero-shot learning, showcasing its robustness even without fine-tuning.
 As expected, the ZF approach performs the worst also under low SNR conditions, delivering sum rates that are approximately 51--416\% lower than the PaPP zero-shot. This sharp decline is mainly because ZF’s simplistic strategy, although useful for interference cancellation, tends to amplify noise, resulting in significantly reduced effectiveness in poor channel conditions.
le performance improvements, particularly at low SNR. 
Notably, at an SNR of 10 dB, fine-tuning PaPP with 4M samples outperforms WMMSE by approximately 5-10\%. This is a ignificant achievement since PaPP also has XY$\times$ lower computatonal complexity than WMMSE., highlighting how the utilization of more diverse training data enhances the model's generalization ability compared to single-site training. Additionally, leveraging a backbone model with limited local fine-tuning reduces the overall training complexity by enabling efficient adaptation to new sites without requiring extensive retraining. This makes the approach computationally efficient while maintaining high performance.\,1873--86, depending on the site

\section{Conclusion} \label{Sec:conclusion}
In this paper, a novel \gls{DL}-based precoding method approach named PaPP is proposed for mMIMO systems to overcome the limitations of traditional methods in terms of complexity and generalizability. By combining WMMSE with DL techniques, we designed a computationally efficient precoder that leverages MLDG and a teacher-student training framework for improved adaptability. Our results demonstrate high sum-rate performance across diverse urban environments while achieving approximately 3.5$\times$ lower computational complexity than alternative generalizable DNN approaches (MAML-CNN).  Furthermore, when deployed in a new site without fine-tuning, PaPP achieves roughly the same performance as a model trained directly on the deployment site. With fine-tuning, PaPP outperforms all other methods. 

\section*{Acknowledgement}
This work was supported by Ericsson - Global Artificial Intelligence Accelerator AI-Hub Canada in Montr\'{e}al and jointly funded by NSERC Alliance Grant 566589-21 (Ericsson, ECCC, Innov\'{E}\'{E}).

\bibliographystyle{IEEEtran}
\bibliography{0_main}

\end{document}